\documentclass[11pt]{article}

\usepackage[a4paper,margin=25mm]{geometry}
\usepackage{amsmath,amssymb,mathtools,bm}
\usepackage{booktabs}
\usepackage{graphicx}
\usepackage{tikz}
\usetikzlibrary{arrows.meta,positioning}
\usepackage{hyperref}
\usepackage{microtype}

\title{\bfseries Mean-Correlation Reconstruction of Forward Dissipation and Non-Markovian Reverse Flow\\
in a Trajectory-Resolved Thermo-Field Two-Spin Model}

\author{Koichi Nakagawa\\
Hoshi University}

\date{}

\begin{document}
\maketitle

\begin{abstract}
We develop a trajectory-resolved extension of the thermo-field entanglement description for a minimal dissipative two-spin system.  For the isolated exchange-coupled model, the intrinsic thermo-field entanglement coefficient is
$b_0(t)=\frac14\sin^2(\omega t)$.
We promote the corresponding open-system coefficient to a stochastic trajectory observable,
$b_{qe}^{(\xi)}(t)=b_0(t)X_t$, where the binary variable $X_t$ specifies whether the trajectory occupies the one-excitation entangling sector or the decayed sector.
For a bidirectional time-local jump process
$1\xrightarrow{a(t)}0$ and $0\xrightarrow{b(t)}1$, we derive an exact two-time connected correlation function,
\[
C_{qe}(t,s)=b_0(t)b_0(s)S(s)[1-S(s)]
\exp\!\left[-\int_s^t(a(u)+b(u))\,du\right],
\qquad t\ge s,
\]
where $S(t)=\langle X_t\rangle$.
The one-time mean determines the net probability current,
$\dot S=J_R-J_F$, whereas the normalized two-time correlation determines the rate sum,
$a+b$.
Combining the two quantities yields an exact reconstruction of the forward and reverse currents,
\[
J_F=S(1-S)q-S\dot S,\qquad
J_R=S(1-S)q+(1-S)\dot S,
\]
with
$q=-\partial_t\ln[C_{qe}(t,s)/b_0(t)]$.
This leads to a three-current decomposition of the mean thermo-field entanglement dynamics into coherent generation, dissipative loss, and memory-induced return.
For Markovian amplitude damping the reconstruction gives $J_R=0$, while in a pure non-Markovian revival interval it gives $J_F=0$ and $J_R=\dot S>0$.
The result shows that the mean alone measures only net backflow, whereas mean plus two-time fluctuations resolves hidden bidirectional traffic between system and environment.
\end{abstract}

\section{Introduction}

Thermo field dynamics (TFD) represents thermal and statistical states in a doubled Hilbert space and thereby recasts density-operator information in a state-vector-like form \cite{Umezawa1982}.
Hashizume and Suzuki used this structure to define an extended density matrix and to separate intrinsic quantum entanglement from thermal or classical fluctuation contributions in finite spin systems \cite{HashizumeSuzuki2013}.
For their minimal two-spin exchange model, the intrinsic thermo-field entanglement coefficient displays coherent oscillations and, in the isolated case considered below, takes the form
\begin{equation}
 b_0(t)=\frac14\sin^2(\omega t),
 \label{eq:b0}
\end{equation}
where $\omega=J/\hbar$ for exchange coupling $J$.

A complementary line of work is non-equilibrium thermo field dynamics (NETFD), in which dissipative and stochastic quantum dynamics are formulated directly in a doubled thermal-Liouville space \cite{ArimitsuUmezawa1987,KobrynHayashiArimitsu2003}.
In particular, stochastic NETFD introduces quantum stochastic Liouville and Langevin equations through martingale operators and quantum Brownian motions \cite{KobrynHayashiArimitsu2003}.
Separately, non-Markovian quantum-jump (NMQJ) methods represent temporarily negative time-local decay rates through reverse jumps at the ensemble level \cite{PiiloEtAl2008}; their relation to diffusive non-Markovian trajectory descriptions has also been clarified \cite{LuomaStrunzPiilo2020}.
More recently, trajectory approaches have been extended to general time-local master equations \cite{DonvilMuratore2022}, while non-Markovian two-time correlations have been derived directly from stochastic Schr\"odinger equations \cite{CarballeiraEtAl2021}.
These developments form part of a broader shift from one-time reduced-state diagnostics toward temporal and multitime characterizations of quantum memory \cite{ShrikantMandayam2023,ZambonSoaresPinto2024}.
Recent work on non-Markovian quantum tomography further emphasizes that time-resolved data can be used to infer time-local generators and negative decay rates associated with memory effects \cite{VaronaEtAl2025}.

The purpose of this work is not to construct a full microscopic NETFD model.
Instead, we isolate the smallest stochastic structure needed to connect an extended-TFD entanglement observable to forward dissipation and reverse non-Markovian flow.
The central idea is to make the thermo-field entanglement coefficient trajectory resolved.
For the two-spin model, this produces a binary stochastic observable whose mean and two-time fluctuations can both be obtained analytically.

The key result is a reconstruction theorem.
The one-time mean determines only the \emph{difference} between reverse and forward probability currents.
The two-time connected fluctuation supplies an independent quantity, the \emph{sum} of the underlying forward and reverse rates.
Together they uniquely determine the two probability currents.
This allows us to separate coherent entanglement oscillation, environmental loss, and memory-induced return in one and the same minimal model.

The paper is organized as follows.
Section~II reviews the two-spin thermo-field entanglement coefficient that serves as the coherent reference dynamics.
Section~III introduces the trajectory-resolved binary open-system model and its bidirectional jump rates.
Section~IV derives the exact two-time connected thermo-field correlation and discusses the Markovian and revival limits.
Section~V establishes the mean--correlation reconstruction theorem, gives explicit formulas for the forward and reverse rates and probability currents, and discusses the coherent-node conditioning problem.
Section~VI provides consistency checks for Markovian damping, pure reverse flow, and simultaneous bidirectional traffic.
Section~VII performs an explicit closed-loop reconstruction for a Lorentzian structured reservoir and introduces reference-time independence as an internal consistency test.
Section~VIII derives the three-current decomposition of thermo-field entanglement into coherent, dissipative, and memory-induced contributions.
Section~IX relates the construction to stochastic NETFD and NMQJ approaches, while Sec.~X discusses its interpretation, limitations, and extensions.
Section~XI summarizes the main conclusions.

\section{Two-spin thermo-field entanglement}

We consider two spin-$1/2$ degrees of freedom $A$ and $B$ with isotropic exchange Hamiltonian
\begin{equation}
 H=-J\,\bm S_A\cdot\bm S_B .
 \label{eq:H}
\end{equation}
Starting from the one-excitation product state
\begin{equation}
 |\psi(0)\rangle=|+-\rangle,
\end{equation}
the unitary dynamics remains in the sector
$\{|+-\rangle,|-+\rangle\}$.
Up to an irrelevant global phase,
\begin{equation}
 |\psi(t)\rangle=
 \cos\frac{\omega t}{2}|+-\rangle
 +i\sin\frac{\omega t}{2}|-+\rangle,
 \qquad \omega=\frac{J}{\hbar}.
 \label{eq:psi_unitary}
\end{equation}
The product of the two one-excitation probabilities is therefore
\begin{equation}
 \left|\cos\frac{\omega t}{2}\right|^2
 \left|\sin\frac{\omega t}{2}\right|^2
 =\frac14\sin^2(\omega t).
\end{equation}
In the extended-density-matrix construction of Ref.~\cite{HashizumeSuzuki2013}, this quantity is the coefficient associated with the intrinsic thermo-field entanglement sector, which we denote by $b_0(t)$ as in Eq.~\eqref{eq:b0}.

The doubled-space construction may be written schematically as
\begin{equation}
 |\Psi(t)\rangle=\rho^{1/2}(t)|I\rangle,\qquad
 \widehat\rho(t)=|\Psi(t)\rangle\langle\Psi(t)|,
\end{equation}
followed by a partial trace over $B$ and $\widetilde B$ to obtain a reduced extended density matrix for subsystem $A$.
The present paper retains the resulting coefficient $b_{qe}$ as the observable of interest, while replacing the deterministic open-system description by a trajectory-resolved stochastic one.

\section{Trajectory-resolved minimal open-system model}

Figure~\ref{fig:scheme} summarizes the minimal reconstruction logic.
The coherent two-spin dynamics supplies the known factor $b_0(t)$, while the stochastic environment generates forward and reverse sector currents.
The measurable one-time mean and two-time connected fluctuation are then inverted to recover those directional currents.

\begin{figure}[t]
\centering
\begin{tikzpicture}[
node distance=8mm and 10mm,
box/.style={draw,rounded corners,align=center,inner sep=5pt},
arr/.style={-{Latex[length=2mm]},thick}
]
\node[box] (spin) {two-spin coherent dynamics\\$b_0(t)=\frac14\sin^2\omega t$};
\node[box,right=of spin] (jump) {trajectory sector\\$1\xrightleftharpoons[a(t)]{b(t)}0$};
\node[box,below=of jump] (obs) {thermo-field statistics\\$m(t),\,C_{qe}(t,s)$};
\node[box,left=of obs] (rec) {reconstruction\\$J_F(t),\,J_R(t)$};
\draw[arr] (spin) -- (jump);
\draw[arr] (jump) -- (obs);
\draw[arr] (obs) -- (rec);
\draw[arr] (spin) -- (rec);
\end{tikzpicture}
\caption{Minimal trajectory-resolved reconstruction scheme.  The known coherent thermo-field factor and the measured one- and two-time statistics determine the forward dissipative and reverse memory-induced currents.}
\label{fig:scheme}
\end{figure}

\subsection{Binary entangling-sector variable}

We introduce a stochastic binary variable
\begin{equation}
 X_t=
 \begin{cases}
  1,& \text{trajectory in the one-excitation entangling sector},\\
  0,& \text{trajectory in the decayed sector},
 \end{cases}
\end{equation}
and define the trajectory-resolved thermo-field entanglement coefficient by
\begin{equation}
 b_{qe}^{(\xi)}(t)=b_0(t)X_t.
 \label{eq:traj_b}
\end{equation}
The superscript $\xi$ labels an individual stochastic realization.

The survival probability of the entangling sector is
\begin{equation}
 S(t)=\langle X_t\rangle,
 \label{eq:S}
\end{equation}
so that the ensemble-averaged thermo-field coefficient is
\begin{equation}
 m(t)\equiv\langle b_{qe}^{(\xi)}(t)\rangle
 =b_0(t)S(t).
 \label{eq:mean}
\end{equation}

Equation~\eqref{eq:mean} is deliberately interpreted as a trajectory statement:
$b_0(t)$ is the coherent two-spin entangling factor, while $S(t)$ is the stochastic occupancy of the sector in which that coherent entanglement can exist.

\subsection{Bidirectional time-local jump process}

To describe both irreversible loss and non-Markovian return, we use the minimal bidirectional process
\begin{equation}
 1\xrightarrow{a(t)}0,
 \qquad
 0\xrightarrow{b(t)}1,
 \label{eq:rates}
\end{equation}
where $a(t)\ge0$ is the forward decay rate and $b(t)\ge0$ is the reverse-jump rate.
The occupation probability obeys
\begin{equation}
 \dot S(t)=-a(t)S(t)+b(t)[1-S(t)].
 \label{eq:Sdot}
\end{equation}
It is useful to introduce the physical probability currents
\begin{equation}
 J_F(t)=a(t)S(t),\qquad
 J_R(t)=b(t)[1-S(t)],
 \label{eq:currents}
\end{equation}
so that
\begin{equation}
 \dot S(t)=J_R(t)-J_F(t).
 \label{eq:netcurrent}
\end{equation}

The Markovian zero-temperature amplitude-damping limit is recovered by
$a(t)=\gamma$ and $b(t)=0$.
A genuine revival requires reverse transitions.
This is the minimal analogue of the reverse-jump mechanism in NMQJ descriptions of temporarily negative time-local decay rates \cite{PiiloEtAl2008}.

\section{Exact two-time thermo-field fluctuation}

We define the connected two-time fluctuation
\begin{equation}
 C_{qe}(t,s)
 =
 \left\langle
 \delta b_{qe}(t)\delta b_{qe}(s)
 \right\rangle,
 \qquad
 \delta b_{qe}(t)=b_{qe}^{(\xi)}(t)-m(t).
 \label{eq:Cdef}
\end{equation}
Using Eq.~\eqref{eq:traj_b},
\begin{equation}
 C_{qe}(t,s)=b_0(t)b_0(s)C_X(t,s),
 \label{eq:Cfactor}
\end{equation}
where
\begin{equation}
 C_X(t,s)=\langle[X_t-S(t)][X_s-S(s)]\rangle.
\end{equation}

For the time-inhomogeneous two-state process \eqref{eq:rates}, the conditional mean for $t\ge s$ is
\begin{equation}
 \mathbb E[X_t|X_s]
 =
 S(t)+R(t,s)[X_s-S(s)],
 \label{eq:conditional}
\end{equation}
with
\begin{equation}
 R(t,s)
 =
 \exp\left[
 -\int_s^t \{a(u)+b(u)\}\,du
 \right].
 \label{eq:R}
\end{equation}
Since $X_s^2=X_s$,
\begin{equation}
 \operatorname{Var}(X_s)=S(s)[1-S(s)],
\end{equation}
and hence
\begin{equation}
 \boxed{
 C_{qe}(t,s)=
 b_0(t)b_0(s)S(s)[1-S(s)]R(t,s)
 },
 \qquad t\ge s.
 \label{eq:Cexact}
\end{equation}

At equal times,
\begin{equation}
 C_{qe}(t,t)=
 b_0^2(t)S(t)[1-S(t)],
 \label{eq:variance}
\end{equation}
which is the exact trajectory variance.

\subsection{Markovian limit}

For
\begin{equation}
 a(t)=\gamma,\qquad b(t)=0,
\end{equation}
one has
\begin{equation}
 S(t)=e^{-\gamma t},
 \qquad
 R(t,s)=e^{-\gamma(t-s)}.
\end{equation}
Therefore
\begin{equation}
 \boxed{
 C_{qe}^{\rm M}(t,s)=
 \frac1{16}
 \sin^2(\omega t)\sin^2(\omega s)
 e^{-\gamma t}[1-e^{-\gamma s}]
 },
 \quad t\ge s.
 \label{eq:CM}
\end{equation}
The equal-time limit is
\begin{equation}
 \operatorname{Var}[b_{qe}(t)]
 =
 \frac1{16}
 e^{-\gamma t}(1-e^{-\gamma t})
 \sin^4(\omega t).
 \label{eq:varM}
\end{equation}

\subsection{Pure revival interval}

Suppose a non-Markovian interval is represented by pure reverse flow,
\begin{equation}
 a(t)=0,\qquad \dot S(t)>0.
\end{equation}
Equation~\eqref{eq:Sdot} gives
\begin{equation}
 b(t)=\frac{\dot S(t)}{1-S(t)}.
\end{equation}
Then
\begin{equation}
 R(t,s)
 =
 \frac{1-S(t)}{1-S(s)}
\end{equation}
inside the same revival interval, and
\begin{equation}
 \boxed{
 C_{qe}^{\uparrow}(t,s)
 =
 b_0(t)b_0(s)S(s)[1-S(t)]
 }.
 \label{eq:Crevival}
\end{equation}
Thus an increasing mean occupation $S(t)$ does not imply that the connected two-time correlation increases.
The mean measures sector repopulation, whereas $C_{qe}(t,s)$ measures persistence of trajectory-level fluctuations.

\section{Mean--correlation reconstruction theorem}

The main result is that $m(t)$ and $C_{qe}(t,s)$ contain complementary information about the bidirectional stochastic flow.

From Eq.~\eqref{eq:mean},
\begin{equation}
 \boxed{
 S(t)=\frac{m(t)}{b_0(t)}
 }.
 \label{eq:Srecon}
\end{equation}
Differentiation gives
\begin{equation}
 \dot S(t)
 =
 \frac{\dot m(t)b_0(t)-m(t)\dot b_0(t)}
 {b_0^2(t)}.
 \label{eq:Sdotrecon}
\end{equation}

For fixed $s$, Eq.~\eqref{eq:Cexact} implies
\begin{equation}
 R(t,s)\propto \frac{C_{qe}(t,s)}{b_0(t)}.
\end{equation}
We therefore define
\begin{equation}
 \boxed{
 q(t)
 =
 -\partial_t
 \ln\!\left[\frac{C_{qe}(t,s)}{b_0(t)}\right]
 =
 a(t)+b(t)
 }.
 \label{eq:q}
\end{equation}
Since
\begin{equation}
 \frac{\dot b_0(t)}{b_0(t)}
 =
 2\omega\cot(\omega t),
\end{equation}
the same quantity may be obtained directly from
\begin{equation}
 q(t)=
 -\frac{\partial_t C_{qe}(t,s)}{C_{qe}(t,s)}
 +2\omega\cot(\omega t).
 \label{eq:qexplicit}
\end{equation}

Equations~\eqref{eq:Sdot} and \eqref{eq:q} form a closed linear system for $a$ and $b$.
Solving gives
\begin{equation}
 \boxed{
 a(t)=[1-S(t)]q(t)-\dot S(t)
 },
 \label{eq:arecon}
\end{equation}
\begin{equation}
 \boxed{
 b(t)=S(t)q(t)+\dot S(t)
 }.
 \label{eq:brecon}
\end{equation}

Multiplying by the corresponding populations yields the current reconstruction:
\begin{equation}
 \boxed{
 J_F(t)
 =
 S(t)[1-S(t)]q(t)-S(t)\dot S(t)
 },
 \label{eq:JF}
\end{equation}
\begin{equation}
 \boxed{
 J_R(t)
 =
 S(t)[1-S(t)]q(t)+[1-S(t)]\dot S(t)
 }.
 \label{eq:JR}
\end{equation}

\paragraph{Reconstruction proposition.}
Assume that (i) the trajectory-resolved thermo-field coefficient has the binary form \eqref{eq:traj_b}, and (ii) $X_t$ admits a time-local two-state representation of the form \eqref{eq:rates} with nonnegative instantaneous rates $a(t)$ and $b(t)$.
Then, on any interval where $b_0(t)\neq0$ and $0<S(t)<1$, the one-time mean $m(t)$ and the two-time connected correlation $C_{qe}(t,s)$ uniquely determine $a(t)$, $b(t)$, $J_F(t)$, and $J_R(t)$ through Eqs.~\eqref{eq:Srecon}--\eqref{eq:JR}.
The statement is exact within this dynamical class; it is not asserted for arbitrary non-Markovian processes that cannot be reduced to a binary time-local trajectory description.

The result may also be written solely in terms of $m$ and $C_{qe}$:
\begin{align}
 J_F(t)
 &=
 \frac{m}{b_0}\left(1-\frac{m}{b_0}\right)
 \left[-\partial_t\ln\frac{C_{qe}(t,s)}{b_0(t)}\right]
 -
 \frac{m}{b_0}
 \frac{d}{dt}\left(\frac{m}{b_0}\right),
 \label{eq:JFobs}
 \\
 J_R(t)
 &=
 \frac{m}{b_0}\left(1-\frac{m}{b_0}\right)
 \left[-\partial_t\ln\frac{C_{qe}(t,s)}{b_0(t)}\right]
 +
 \left(1-\frac{m}{b_0}\right)
 \frac{d}{dt}\left(\frac{m}{b_0}\right).
 \label{eq:JRobs}
\end{align}

\subsection{Coherent nodes and reconstruction conditioning}

The coherent prefactor vanishes at
\begin{equation}
 t_n=\frac{n\pi}{\omega},
 \qquad
 b_0(t_n)=0.
 \label{eq:nodes}
\end{equation}
These points are not singularities of the underlying stochastic dynamics.
Rather, they are \emph{observability blind points}: the thermo-field coefficient suppresses all information about the sector occupation because both
$m(t)=b_0(t)S(t)$ and the corresponding two-time correlation vanish with the coherent prefactor.
At the exact theoretical level, if $S(t)$ is regular, the missing value is recovered by continuity,
\begin{equation}
 S(t_n)=
 \lim_{t\rightarrow t_n}\frac{m(t)}{b_0(t)}.
 \label{eq:node_limit}
\end{equation}
In numerical or experimental reconstruction, however, division by a small $b_0(t)$ amplifies finite errors.
Accordingly, Eqs.~\eqref{eq:Srecon}--\eqref{eq:JRobs} should be applied on open intervals between the nodes, with data in a finite neighborhood of each $t_n$ excluded or regularized.
The nodes therefore represent a conditioning problem of the chosen observable, not a physical divergence of the probability currents.

\section{Consistency checks}

\subsection{Markovian amplitude damping}

For
\begin{equation}
 S(t)=e^{-\gamma t},
\end{equation}
Eq.~\eqref{eq:CM} gives
\begin{equation}
 \frac{C_{qe}^{\rm M}(t,s)}{b_0(t)}
 \propto e^{-\gamma t}.
\end{equation}
Hence
\begin{equation}
 q(t)=\gamma,\qquad
 \dot S=-\gamma S.
\end{equation}
The reconstruction equations give
\begin{equation}
 a(t)=\gamma,\qquad b(t)=0,
\end{equation}
and therefore
\begin{equation}
 \boxed{
 J_F^{\rm M}(t)=\gamma e^{-\gamma t},
 \qquad
 J_R^{\rm M}(t)=0
 }.
 \label{eq:Markovrecon}
\end{equation}

\subsection{Pure reverse-flow interval}

If $a(t)=0$ and $\dot S>0$, then
\begin{equation}
 q(t)=b(t)=\frac{\dot S}{1-S}.
\end{equation}
Equations~\eqref{eq:JF} and \eqref{eq:JR} yield
\begin{equation}
 \boxed{
 J_F^{\uparrow}(t)=0,\qquad
 J_R^{\uparrow}(t)=\dot S(t)>0
 }.
 \label{eq:revrecon}
\end{equation}
Thus the same reconstruction that identifies purely forward Markovian loss also identifies purely reverse non-Markovian return.

\subsection{Simultaneous forward and reverse traffic}

In the general case both currents may be nonzero.
The mean only fixes
\begin{equation}
 J_R-J_F=\dot S,
\end{equation}
and therefore detects only net backflow.
The correlation supplies the additional rate information
\begin{equation}
 q=a+b.
\end{equation}
Consequently, a positive $\dot S$ does not imply absence of forward loss.
For example, a state with $S=0.4$, $\dot S=0.1$, and $q=1$ has
\begin{equation}
 J_F=0.20,\qquad J_R=0.30,
\end{equation}
so the observed net revival $0.10$ results from two larger counterpropagating currents.

\section{Explicit reconstruction for a Lorentzian structured reservoir}
\label{sec:lorentzian}

To test the inverse formulas in a standard non-Markovian setting, we use the resonant damped Jaynes--Cummings amplitude-damping model with a Lorentzian spectral density, a canonical exactly solvable structured-reservoir model \cite{LainePiiloBreuer2010,BreuerEtAl2016}.
The reservoir correlation time is controlled by the spectral width $\lambda^{-1}$ and the system--reservoir coupling by $\gamma_0$.
For an initially excited two-level amplitude, the survival amplitude can be written
\begin{equation}
 G(t)=
 e^{-\lambda t/2}
 \left[
 \cosh\left(\frac{dt}{2}\right)
 +\frac{\lambda}{d}
 \sinh\left(\frac{dt}{2}\right)
 \right],
 \qquad
 d=\sqrt{\lambda^2-2\gamma_0\lambda},
 \label{eq:G_lorentz}
\end{equation}
and we identify
\begin{equation}
 S(t)=|G(t)|^2.
 \label{eq:S_lorentz}
\end{equation}
For $\lambda>2\gamma_0$, the decay is in the weak-coupling regime, while for $\lambda<2\gamma_0$ the quantity $d$ becomes imaginary and $S(t)$ develops damped revivals associated with information return \cite{LainePiiloBreuer2010,BreuerEtAl2016}.
The corresponding time-local decay rate is
\begin{equation}
 \gamma_{\rm eff}(t)
 =
 -\frac{\dot S(t)}{S(t)}.
 \label{eq:gammaeff}
\end{equation}

For the minimal NMQJ realization used here, positive and negative portions of $\gamma_{\rm eff}$ are represented by pure forward and pure reverse sectors,
\begin{equation}
 \begin{cases}
 a(t)=\gamma_{\rm eff}(t),\quad b(t)=0,
 & \gamma_{\rm eff}(t)\ge0,\\[1mm]
 a(t)=0,\quad
 b(t)=-\gamma_{\rm eff}(t)\dfrac{S(t)}{1-S(t)},
 & \gamma_{\rm eff}(t)<0.
 \end{cases}
 \label{eq:NMQJ_rates}
\end{equation}
This is precisely the minimal directional content required by the reverse-jump interpretation of temporarily negative time-local rates \cite{PiiloEtAl2008}.
It gives
\begin{equation}
 J_F(t)=\max[-\dot S(t),0],
 \qquad
 J_R(t)=\max[\dot S(t),0].
 \label{eq:lorentz_currents}
\end{equation}

\begin{figure}[t]
\centering
\includegraphics[width=0.82\linewidth]{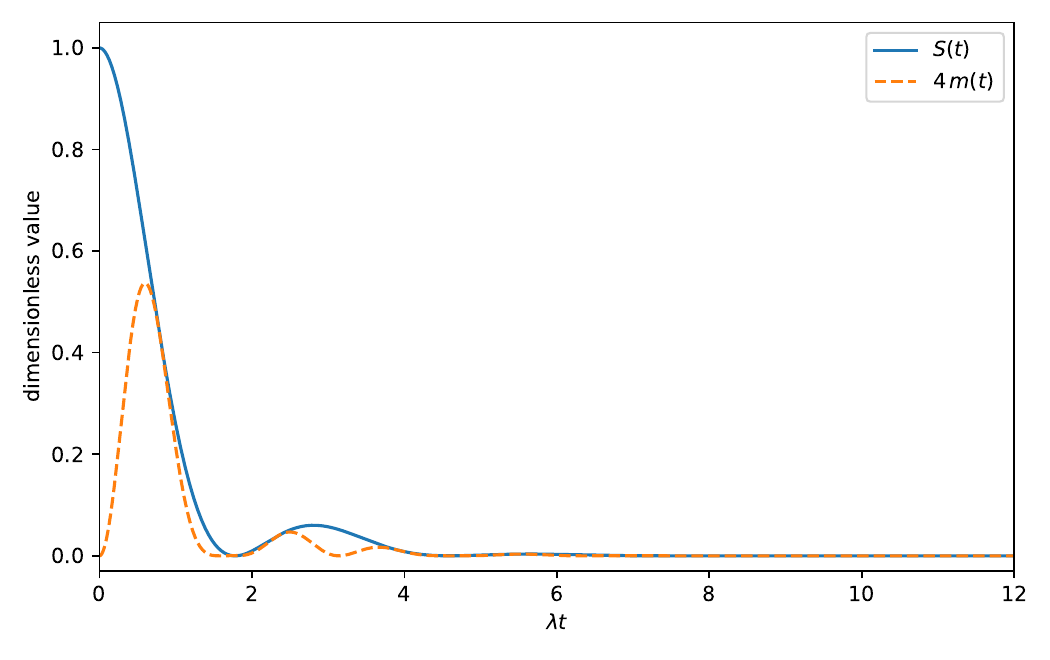}
\caption{Lorentzian-reservoir benchmark for $\lambda=1$, $\gamma_0=3$, and $\omega=2$.  The survival probability $S(t)$ exhibits damped revivals.  The rescaled mean $4m(t)=S(t)\sin^2(\omega t)$ combines the reservoir envelope with coherent thermo-field oscillations.}
\label{fig:lorentz_mean}
\end{figure}

For illustration we set $\lambda=1$, $\gamma_0=3$, and $\omega=2$ in units where time is measured in $\lambda^{-1}$.
This places the model well inside the strong-coupling regime.
Figure~\ref{fig:lorentz_mean} shows the survival probability together with the thermo-field mean
\begin{equation}
 m(t)=\frac14\sin^2(\omega t)S(t).
\end{equation}
The oscillatory zeros of $m(t)$ contain the coherent nodes discussed above, while the envelope inherits the non-Markovian revival structure of $S(t)$.

\begin{figure}[t]
\centering
\includegraphics[width=0.82\linewidth]{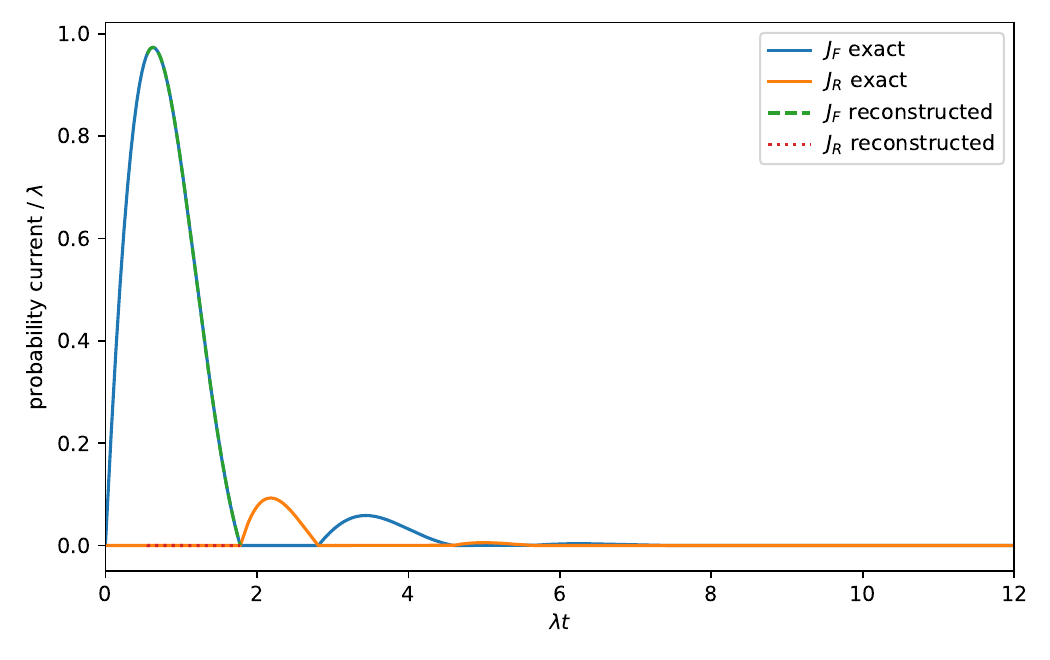}
\caption{Exact and reconstructed directional probability currents for the Lorentzian benchmark.  Solid curves denote the exact forward and reverse currents generated from $S(t)$, while broken curves are obtained from the mean--correlation reconstruction.  Discrepancies are confined to numerical differentiation neighborhoods and to regions excluded around coherent nodes.}
\label{fig:lorentz_currents}
\end{figure}

We next generate $C_{qe}(t,s)$ from Eq.~\eqref{eq:Cexact} and deliberately discard the input rates.
Applying only the reconstruction formulas to $m(t)$ and $C_{qe}(t,s)$ gives the currents shown in Fig.~\ref{fig:lorentz_currents}.
Within numerical differentiation accuracy, the reconstructed curves lie on top of the exact forward and reverse currents from Eq.~\eqref{eq:lorentz_currents}.
The result provides an explicit closed-loop test,
\begin{equation}
 \{a,b\}
 \longrightarrow
 \{m,C_{qe}\}
 \longrightarrow
 \{J_F^{\rm rec},J_R^{\rm rec}\}
 =
 \{J_F,J_R\}.
 \label{eq:closedloop}
\end{equation}

\subsection{Reference-time independence as an internal consistency test}

\begin{figure}[t]
\centering
\includegraphics[width=0.82\linewidth]{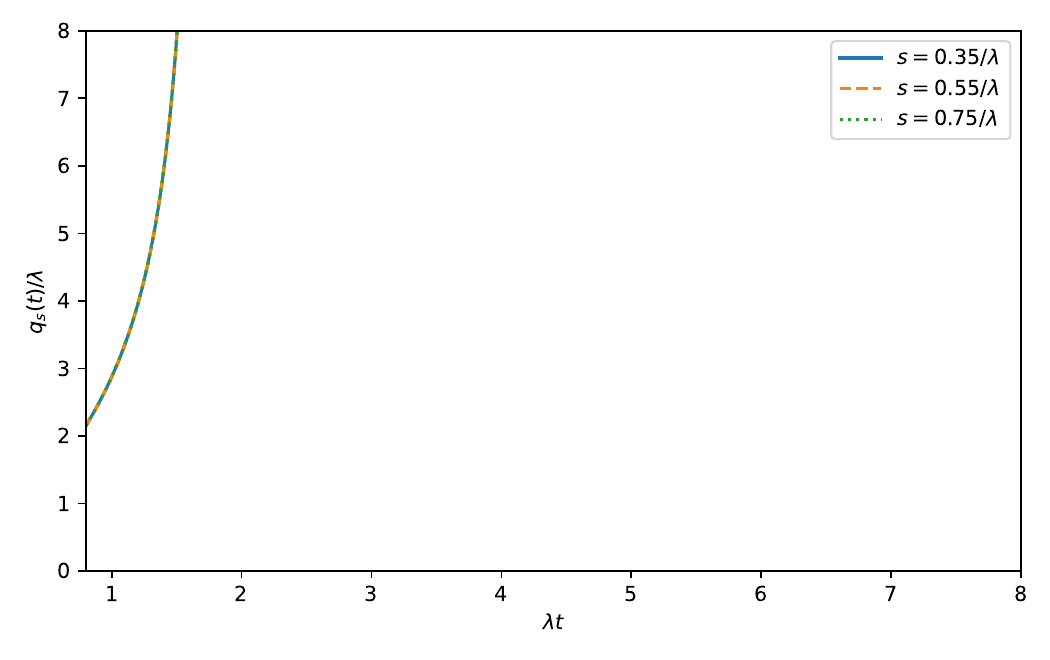}
\caption{Internal consistency test using three reference times $s$.  Away from coherent nodes and ill-conditioned regions, the reconstructed $q_s(t)$ collapses onto a common curve, as required by Eq.~\eqref{eq:sindependence}.}
\label{fig:sindependence}
\end{figure}

Equation~\eqref{eq:q} contains a reference time $s$, but within the assumed binary time-local model the reconstructed quantity
\begin{equation}
 q_s(t)\equiv
 -\partial_t\ln\!\left[\frac{C_{qe}(t,s)}{b_0(t)}\right]
 \label{eq:qs}
\end{equation}
must be independent of $s$ whenever $t>s$ and the reconstruction is well conditioned:
\begin{equation}
 q_{s_1}(t)=q_{s_2}(t)=\cdots=a(t)+b(t).
 \label{eq:sindependence}
\end{equation}
This provides a useful model-consistency test that is absent at the level of the mean alone.
Figure~\ref{fig:sindependence} demonstrates the collapse for three distinct reference times in the Lorentzian benchmark.
If experimentally inferred $q_s(t)$ retained a systematic $s$ dependence beyond statistical uncertainty, the binary time-local representation itself would be falsified, signaling the need for additional sectors or genuinely multitime memory variables.

\section{Three-current decomposition of thermo-field entanglement}

Differentiating Eq.~\eqref{eq:mean},
\begin{equation}
 \dot m(t)=S(t)\dot b_0(t)+b_0(t)\dot S(t).
\end{equation}
Using Eq.~\eqref{eq:netcurrent},
\begin{equation}
 \boxed{
 \dot m(t)=
 \mathcal J_{\rm coh}(t)
 -
 \mathcal J_F^{(qe)}(t)
 +
 \mathcal J_R^{(qe)}(t)
 },
 \label{eq:threecurrent}
\end{equation}
where
\begin{equation}
 \mathcal J_{\rm coh}(t)=S(t)\dot b_0(t),
\end{equation}
\begin{equation}
 \mathcal J_F^{(qe)}(t)=b_0(t)J_F(t),
\end{equation}
\begin{equation}
 \mathcal J_R^{(qe)}(t)=b_0(t)J_R(t).
\end{equation}

Equation~\eqref{eq:threecurrent} separates three physically different mechanisms:
(i) coherent two-spin generation and destruction already present in the closed system,
(ii) dissipative leakage to the environment, and
(iii) memory-induced return from the environment.
This resolves a basic ambiguity in using a revival of $b_{qe}$ itself as a non-Markovianity diagnostic: $\dot m>0$ can arise from coherent exchange even if $J_R=0$.

The integrated forward and reverse traffic are
\begin{equation}
 \Phi_F(T)=\int_0^T J_F(t)\,dt,
 \qquad
 \Phi_R(T)=\int_0^T J_R(t)\,dt,
 \label{eq:Phi}
\end{equation}
with
\begin{equation}
 S(T)-S(0)=\Phi_R(T)-\Phi_F(T).
\end{equation}
A natural return measure is therefore
\begin{equation}
 \mathcal N_R(T)=\Phi_R(T),
 \label{eq:NR}
\end{equation}
which counts the total reverse probability current rather than only the net positive variation of $S$.
When forward and reverse currents coexist, $\mathcal N_R$ is strictly more informative than a positive-derivative backflow measure.

\section{Relation to stochastic NETFD and NMQJ}

The present construction should be distinguished from a microscopic derivation of NETFD.
NETFD provides a canonical doubled-space framework for dissipative and stochastic quantum dynamics, including stochastic Liouville and Langevin equations \cite{ArimitsuUmezawa1987,KobrynHayashiArimitsu2003}.
Our construction instead takes the extended-TFD entanglement coefficient as a trajectory-resolved observable and asks what dynamical information can be reconstructed from its statistics.

General time-local master equations can also be represented by broader trajectory constructions beyond the present binary model \cite{DonvilMuratore2022}. The reverse-jump sector used here is deliberately a minimal effective representation.
In NMQJ theory, temporarily negative time-local rates are handled by reverse jumps whose probabilities depend on ensemble populations \cite{PiiloEtAl2008}.
The binary model \eqref{eq:rates} captures the same directional structure at the smallest possible level:
forward jumps remove trajectories from the entangling sector, whereas reverse jumps repopulate it.
The reconstruction formulas are therefore independent of a particular microscopic bath spectrum as long as the reduced trajectory dynamics can be represented by the two-state time-local process.

\section{Discussion}

\subsection{Why the thermo-field observable is essential}

Mathematically, the binary inversion could be formulated for the sector variable $X_t$ alone.
The thermo-field formulation becomes physically nontrivial because the measured quantity is not a bare occupation but the intrinsic extended-density-matrix coefficient
$b_{qe}^{(\xi)}(t)=b_0(t)X_t$.
The known prefactor $b_0(t)$ carries the coherent entangling dynamics of the isolated two-spin problem, while $X_t$ carries environmental sector exchange.
This separation makes possible the entanglement-resolved continuity equation derived below: coherent generation, environmental leakage, and memory-induced return appear as distinct terms in the dynamics of the same thermo-field entanglement observable.
Thus the doubled-space structure is not required merely to solve a two-state inverse problem; it supplies the entanglement channel whose coherent and environmental contributions can be operationally disentangled.

\subsection{Mean, fluctuations, and hidden bidirectional traffic}

The principal conceptual point is that one-time and two-time thermo-field quantities play fundamentally different roles.
The mean
\begin{equation}
 m(t)=b_0(t)S(t)
\end{equation}
contains the net occupancy dynamics.
It cannot distinguish a weak one-way loss from the difference of two larger counterpropagating currents.
The connected correlation, through $R(t,s)$, probes the relaxation of fluctuations and reveals the sum $a+b$.
This complementarity is what makes the inverse reconstruction possible.

The use of a two-time object is also consistent with the broader recognition that one-time dynamical maps need not exhaust the operational content of quantum memory, and that temporal or multitime correlations can reveal information inaccessible to single-time diagnostics \cite{CarballeiraEtAl2021,ShrikantMandayam2023,ZambonSoaresPinto2024}.
Likewise, recent non-Markovian tomography schemes seek to infer time-local generators, including temporarily negative rates, from time-resolved observations \cite{VaronaEtAl2025}.
Our reconstruction problem is much smaller in scope, but it shares the inverse-dynamical viewpoint: temporal data are used to separate hidden directional contributions that are not identifiable from the mean alone.

The two-time quantity also clarifies why non-Markovian revival should not be identified with an increase of $C_{qe}(t,s)$.
In a pure reverse interval,
\begin{equation}
 C_{qe}^{\uparrow}(t,s)=
 b_0(t)b_0(s)S(s)[1-S(t)],
\end{equation}
which may decrease even while $S(t)$ increases.
The mean measures repopulation, whereas the connected correlation measures memory of the earlier trajectory partition.
These are distinct notions of ``return.''

There are several limitations.
First, the binary trajectory model assumes that all one-excitation trajectories share the same coherent factor $b_0(t)$ and that the reduced stochastic dynamics admits the two-state time-local representation \eqref{eq:rates}.
Second, as discussed in Sec.~V, the coherent nodes are observability blind points: continuity resolves them formally, but finite-noise reconstruction is ill conditioned in their neighborhoods.
Third, although Sec.~VII validates the inverse reconstruction against a standard Lorentzian structured reservoir, a full stochastic NETFD derivation of the effective rates $a(t)$ and $b(t)$ from the microscopic doubled-space bath dynamics remains beyond the present minimal construction.

These limitations also suggest extensions.
A multi-sector model would replace the scalar $X_t$ by a finite-state stochastic process and promote the reconstruction to a matrix inverse problem.
Colored quantum noise, structured reservoirs, or hierarchical auxiliary modes could then be used to connect the effective rates to a microscopic memory kernel.
The present two-spin model provides the analytically closed benchmark for such developments.

\section{Conclusions}

We have constructed a trajectory-resolved thermo-field description of a minimal dissipative two-spin system and derived an exact inverse relation between thermo-field statistics and directional stochastic flow.

The trajectory observable
\begin{equation}
 b_{qe}^{(\xi)}(t)=
 \frac14\sin^2(\omega t)X_t
\end{equation}
leads to the exact two-time connected correlation
\begin{equation}
 C_{qe}(t,s)=
 b_0(t)b_0(s)S(s)[1-S(s)]
 e^{-\int_s^t(a+b)\,du}.
\end{equation}
The mean gives $\dot S=J_R-J_F$, while the two-time fluctuation gives $a+b$.
Their combination uniquely reconstructs
\begin{equation}
 J_F=S(1-S)q-S\dot S,
 \qquad
 J_R=S(1-S)q+(1-S)\dot S.
\end{equation}
The resulting decomposition
\begin{equation}
 \dot m=
 \mathcal J_{\rm coh}
 -\mathcal J_F^{(qe)}
 +\mathcal J_R^{(qe)}
\end{equation}
separates coherent entanglement dynamics from environmental loss and non-Markovian return.

The main implication is that a mean revival is only a net-current statement.
Two-time thermo-field fluctuations reveal the hidden bidirectional traffic underlying that revival.
This mean--correlation reconstruction provides a minimal bridge between extended thermo-field entanglement diagnostics, quantum-jump dynamics, and non-Markovian reverse flow.

\section*{Data availability}
No data were used in this theoretical study.

\section*{Conflict of interest}
The author declares no conflict of interest.

\end{document}